\documentclass[aps,prl,twocolumn,superscriptaddress,showpacs]{revtex4}
\usepackage{longtable}
\usepackage[dvips]{graphicx}

\begin{document}


\title{Superconductivity in iron telluride thin films under tensile stress}


\author{Y.~Han}
\affiliation{National Laboratory for Superconductivity, Institute of
Physics and Beijing National Laboratory for Condensed Matter
Physics, Chinese Academy of Sciences, Beijing 100190, China}
\author{W.Y.~Li}
\affiliation{National Laboratory for Superconductivity, Institute of
Physics and Beijing National Laboratory for Condensed Matter
Physics, Chinese Academy of Sciences, Beijing 100190, China}
\author{L.X.~Cao}
\email{lxcao@aphy.iphy.ac.cn}
\affiliation{National Laboratory for Superconductivity, Institute of
Physics and Beijing National Laboratory for Condensed Matter
Physics, Chinese Academy of Sciences, Beijing 100190, China}
\author{X.Y.~Wang}
\affiliation{Institute of Semiconductors, Chinese Academy of Sciences, Beijing 100083, China}
\author{B.~Xu}
\affiliation{National Laboratory for Superconductivity, Institute of
Physics and Beijing National Laboratory for Condensed Matter
Physics, Chinese Academy of Sciences, Beijing 100190, China}
\author{B.R.~Zhao}
\affiliation{National Laboratory for Superconductivity, Institute of
Physics and Beijing National Laboratory for Condensed Matter
Physics, Chinese Academy of Sciences, Beijing 100190, China}
\author{Y.Q.~Guo}
\affiliation{School of Energy and Power Engineering, North China
Electric Power University, Beijing 102206, China}
\author{J.L.~Yang}
\affiliation{Institute of Semiconductors, Chinese Academy of Sciences, Beijing 100083, China}



\date{\today}

\begin{abstract}
By realizing in thin films a tensile stress state, superconductivity of 13 K
was introduced into FeTe, an non-superconducting parent compound of the iron
pnictides and chalcogenides, with transition temperature higher than
that of its superconducting isostructural counterpart FeSe. For
these tensile stressed films, the superconductivity is accompanied
by the softening of the first-order magnetic and structural phase
transition; and also, the in-plane extension and out-of-plane
contraction are universal in all FeTe films independent of sign of
lattice mismatch, either positive or negative. Moreover, the
correlations were found exist between the transition temperatures
and the tetrahedra bond angles in these thin films.
\end{abstract}


\maketitle

There is considerable interest in
promoting transition temperature (T$_{c}$) \cite{Locquet} and even
introducing superconductivity by realizing in thin films a
high-pressure state, i.e., an effect of the stress tensor. Stress in
thin film is, for specified directions, the force per unit length
the substrate exerted across the interface on the elastically
deformed film. Although an in-plane extension of the film is not
forbidden in nature, stress effect is generally believed to be
in-plane contraction analogous to application of hydrostatic
pressure on bulk materials \cite{Locquet}. Consequently, tensile
stress in films is usually believed to be irrelevant to such purpose and
therefore is rarely studied.

Very recently, the discovery of superconductivity in iron pnictides
\cite{KamiharaJACS} and chalcogenides \cite{HsuPNAS} has triggered
tremendous efforts to search for new superconductor materials and to
raise their T$_{c}$ by chemical doping \cite{Rotter,Tapp,ChenXHnature,RenZAcpl}
or by external pressure \cite{TakahashiH,Torikachvili,MizuguchiAPL,Medvedev}.
A corrugated layer comprising Fe and pnictogens (Pn =
P, As) or chalcogens (Ch = Se, Te) incorporates with different
interlayers leading to four structural families, among which binary
iron chalcogenide FeSe and FeTe as well as their solid solution
FeSe$_{1-x}$Te$_{x}$ possess the simplest crystal structure with only the FeCh
layers stacking one by another.
In comparison with chemical
doping which usually changes physical parameters in many different
ways, hydrostatic pressure experiment can provide systematic study
on salient physics, and therefore it is widely applied to study
phase transitions and to raise T$_{c}$ of the iron pnictides \cite{TakahashiH,Torikachvili} and
chalcogenides \cite{MizuguchiAPL,Medvedev}, including making parent compounds
superconducting \cite{Torikachvili}.

We report the superconductivity at 13 K in FeTe
which is in the form of thin films and under the tensile stress,
although bulk crystals are not superconducting at ambient pressure
\cite{LiSLprb,BaoW,ChenGF} or under high pressure \cite{MizuguchiPhysicaC,ZhangC}.
The intriguing fact is, superconductivity
appears when the first-order magnetic and structural phase
transition softens, and when the Fe-Te-Fe bond angles become larger.
Our demonstration of
superconductivity introduced by extension of lattice realized via
interfacial stress paves the way for higher transition temperatures
in iron chalcogenides by fine tuning the crystal structure through
chemical doping and for better understanding of superconductivity
mechanisms in this category of materials.

Over 100 FeTe films were pulsed laser deposited with chamber
base pressure of $4\times10^{-5}$ Pa and at $\sim 540\,^\circ \mathrm{C}$
under environment better than $2\times10^{-4}$ Pa on (001)-oriented
$4\times 5\times0.5\,$mm$^3$ (LaAlO$_3$)$_{0.3}$(SrAl$_{0.5}$Ta$_{0.5}$O$_3$)$_{0.7}$ (LSAT), MgO, SrTiO$_3$, and
LaAlO$_{3}$ substrates, respectively \cite{HanY}.
All films are superconducting. An XeCl excimer laser with a repetition
rate of 4 Hz and power density of 100 mJ$\,$mm$^{-2}$ was used, giving a
deposition rate of ~0.05 nm per laser pulse. Targets with nominal
composition of FeTe$_{1+x}$ ($x=0,\,0.2,\,0.4$) were vacuum sintered
twice at $\sim 600\,^\circ \mathrm{C}$ for 24 h \cite{HanY} with excess Te up to 40$\,$\%
to compensate volatile Te losses in FeTe films.

Although FeTe films deposited from a FeTe target
without excess Te content are superconducting, epitaxial, and single-phased as revealed
by X-ray diffraction (XRD), some Fe-rich precipitates of $\sim500$ nm in size are
found on surfaces of such films, 24 in total. Scanning electron
microscopy (SEM) and energy dispersive analysis of X-ray (EDAX)
give a Fe : Te ratio of $\sim8.9 : 1$ with electron beam focused
on the precipitates.

\begin{figure}
\includegraphics[angle=0,width=75mm]{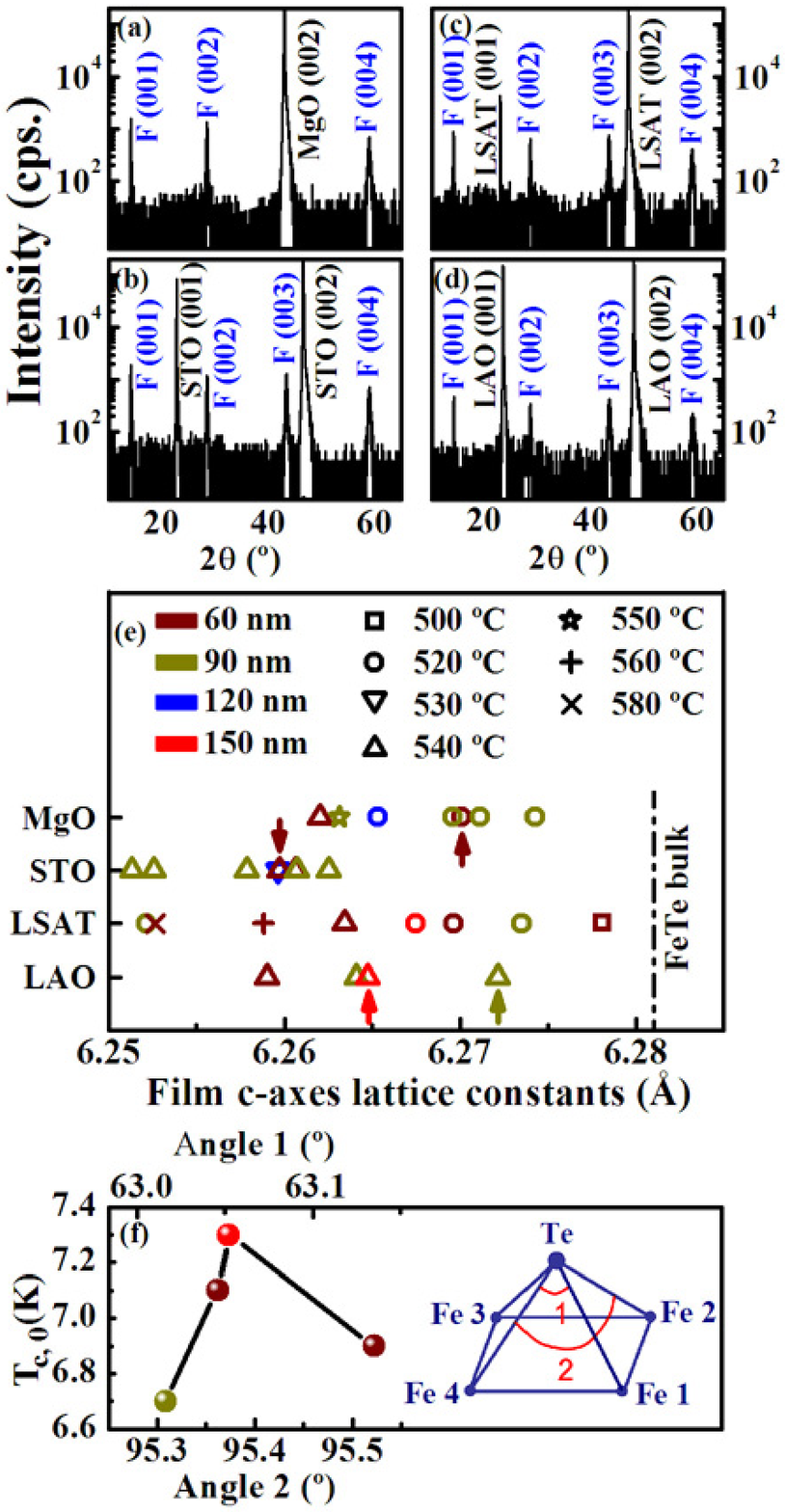}%
\caption{(color). X-ray diffraction spectra measured at room temperature and related analysis. (a)-(d) XRD
spectra of films on MgO, SrTiO$_{3}$, LSAT and LaAlO$_{3}$, respectively, with F denoting film. (e)
Summary of c-axes lattice constants for 32 FeTe films on 4 different
substrates. Arrows indicate 4 samples the high-resolution XRD of
which were performed to deduce the Fe-Te-Fe angles. (f) Fe-Te-Fe bond
angle dependencies of zero resistance transition temperatures T$_{c,\,0}$
for 4 FeTe films indicated by arrows in (e). Inset shows schematically
the definition of the angle 1 and angle 2, which are Fe4-Te-Fe1 and Fe4-Te-Fe2, respectively. \label{fig1}}
\end{figure}

All films given in this paper, 86 in total, are deposited from the targets
with nominal composition of FeTe$_{1.4}$. As shown in
Fig. \ref{fig1}(a)-(d), they are single-phased and (001)-oriented. The
c-axes lattice constants of the FeTe films were calibrated
by those of substrate single-crystals, as given in Fig. \ref{fig1}(e)
for 32 films with thicknesses of 60, 90, 120, and 150 nm
deposited on 4 different substrates at temperatures ranging
from $500\,^\circ \mathrm{C}$ to $580\,^\circ \mathrm{C}$.
Thickness dependencies of the zero resistance transition
temperature T$_{c,\,0}$ reveal the maxima in 90 nm films,
which suggest the critical thicknesses for pseudomorphic
growth of these in-plane stretched films are $\sim90$ nm \cite{CaoLXprb}.
High-resolution XRD was performed on 4 out of 32 samples shown
in Fig. \ref{fig1}(e) to obtain their a-axis lattice constants, from
which the Fe-Te-Fe bond angles can be estimated given in Fig. \ref{fig1}(f).
Meanwhile the Rietveld refinement was performed on a FeTe powder
sample at 300$\,$K. Minor amount of FeTe$_2$ ($\sim7\,$\%) was found
coexist with Fe$_{1.08}$Te phase, the a, c, Te z-coordination, angle 1, and angle 2 of which are
0.38214(3)$\,$nm, 0.62875(3)$\,$nm, 0.2803(2), 62.64$\,^\circ$, and 94.63$\,^\circ$, respectively. This refinement result is in consistent with that given in Ref. \cite{MizuguchiPhysicaC}.

One striking character is that the superconducting FeTe films
change dramatically in the Fe-Te-Fe angles compared to the
non-superconducting Fe$_{1+\delta}$Te bulk samples. The
increments of the angle 1 and angle 2 are $\sim0.4\,^\circ$
and $\sim0.75\,^\circ$, respectively (Fig. \ref{fig1}(f)).
The bond angles dependencies of T$_{c,\,0}$ are given in Fig. \ref{fig1}(f).

Since a-axis lattice constant of the FeTe bulk material is smaller
than those of MgO, SrTiO$_3$, and LSAT, but larger than that of the
LaAlO$_3$, the fact shown in Fig. \ref{fig1}(e) that all c-axes lattice constants
of the films are smaller than that of the FeTe bulk is peculiar. The
contradiction to the expectation that the films on LaAlO$_{3}$ should be
compressed in-plane and therefore stretched out-of-plane suggests
that FeTe is quite unique in properties. This may include at least 3
aspects: (1) FeTe intends to expand its lattice in-plane regardless
of sign of the lattice mismatch, which is possible since epitaxy of
thin films is a kind of low dimensional phenomenon providing more
freedom for lattice to adjust itself. (2) FeTe can easily shrink its
lattice out-of-plane in case needed. (3) the critical thickness of
~90 nm implies smaller elastic modulus, i.e., FeTe being a softer
material. The above hypotheses receive strong supports from the
recent hydrostatic experimental results on FeTe \cite{MizuguchiPhysicaC,ZhangC} and on FeSe
\cite{Millican,Braithwaite,MizuguchiAPL,Medvedev}, from which the Se-Se bonds between the adjacent
FeSe layers are van der Waals force; and FeTe and FeSe are difficult
to be compressed in-plane, easier to be compressed out-of-plane, and
softer with an elastic modulus as small as only $\sim30$ GPa for FeSe
\cite{Millican}.

\begin{figure}
\includegraphics[angle=0,width=75mm]{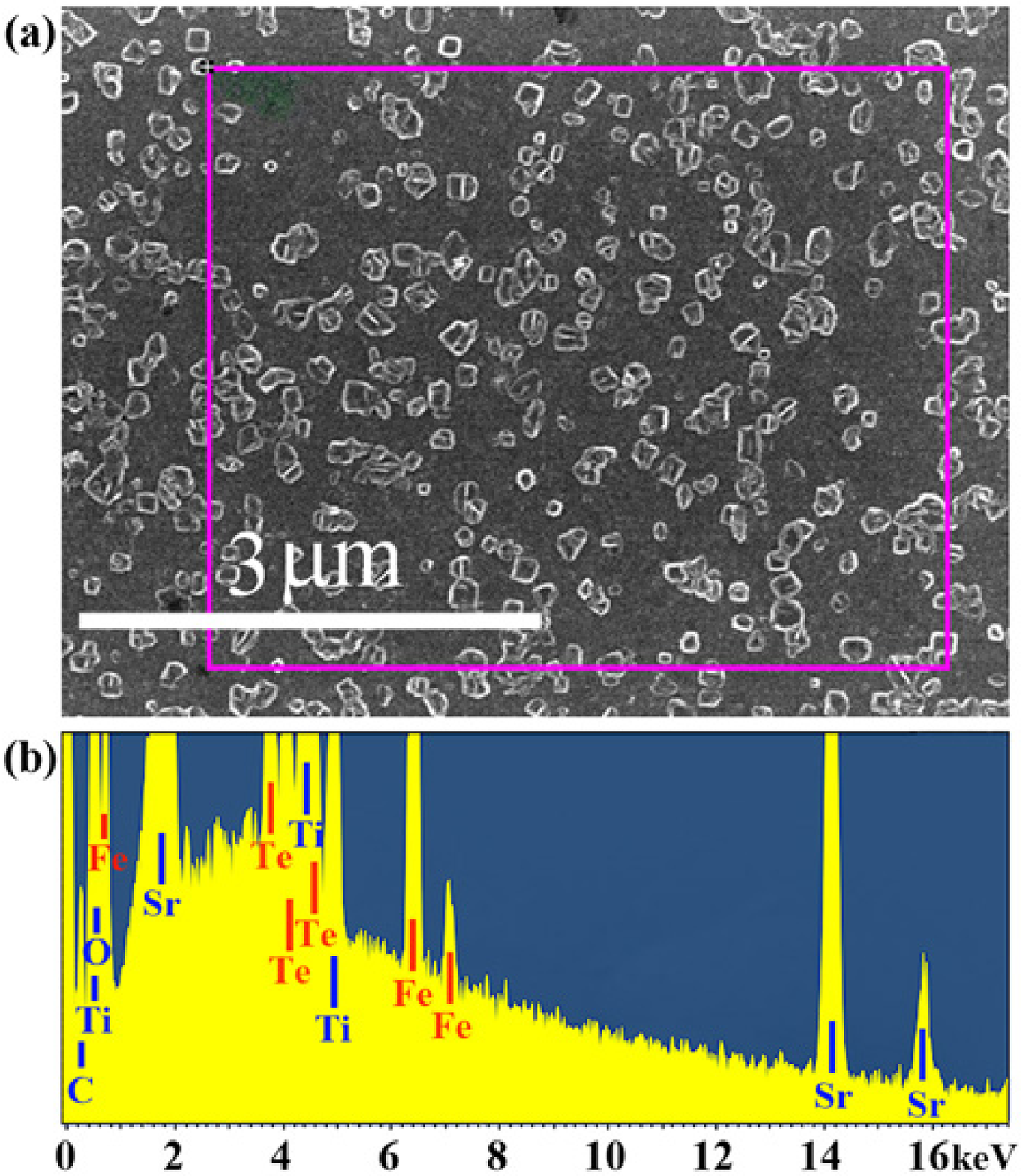}%
\caption{(color). Scanning electron microscopy and energy dispersive analysis
of X-ray. (a) SEM photograph of a FeTe film on SrTiO$_{3}$. (b) EDAX of
the surface area within the frame shown in (a), giving a Fe : Te ratio
of 1.08 : 1. \label{fig2}}
\end{figure}

The FeTe films on different substrates were observed by the SEM,
atomic force microscopy, and transmission electron microscopy. The
SEM and EDAX results of a FeTe film on SrTiO$_{3}$ (Fig. \ref{fig2}(a),(b)) reveal
that the film has precipitates of $\sim150$ nm in size on the surface.
The EDAX were performed not only on the big area (Fig. \ref{fig2}(b)), but also
on the precipitates and on the areas without precipitate. These
analyses lead to 2 results: (1) the Fe : Te ratios ranging from 1.04
to 1.10 over 1, (2) no Se contaminations in the FeTe films.

\begin{figure}
\includegraphics[angle=0,width=75mm]{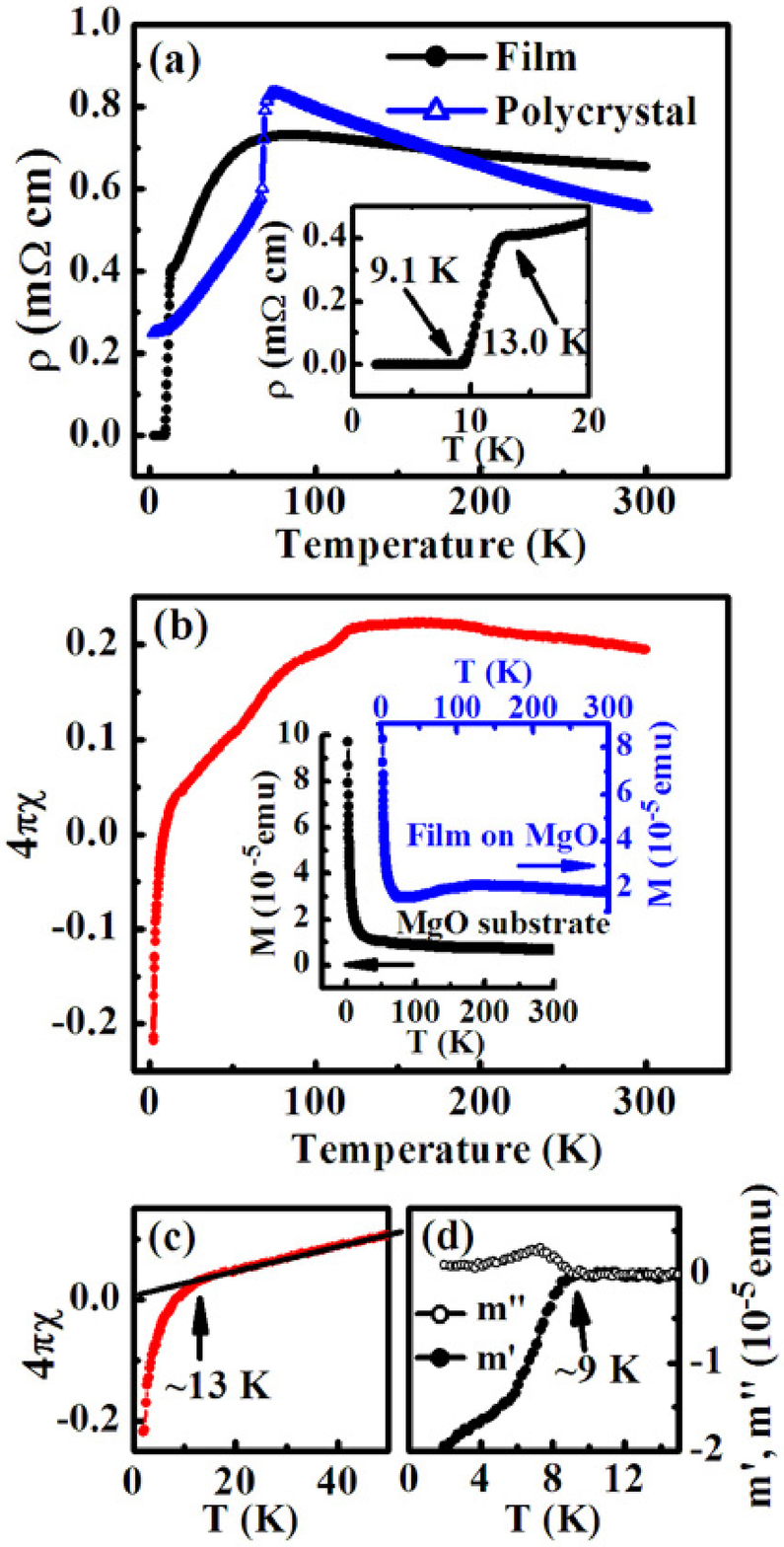}%
\caption{(color). Resistivity and magnetization versus temperature
measurements for a FeTe film on MgO substrate measured from 300$\,$K down to 2$\,$K.
(a) 4 probe resistivity
measurement result for a film and a FeTe bulk crystal given as a
reference. Inset gives the enlargement of part of that of the film.
The first-order phase transition at 70 K can be seen clearly in the
FeTe bulk crystal. (b) Subtraction result of the dc magnetization of
the FeTe film, with magnetic field perpendicular to the film surface plane.
Inset shows the original dc magnetization measurement
results for the FeTe film on MgO, and for the MgO substrate.
Measurements were performed under 500 Oe and with the field
perpendicular to the film surface. (c) Enlargement of part of the
subtraction result in (b). (d) The ac susceptibility measurement
result. \label{fig3}}
\end{figure}

Figure \ref{fig3} reproduces the temperature dependencies of the resistivity,
dc magnetization, and ac susceptibility of a FeTe film on MgO substrate.
The starting transition temperature T$_{c,\,onset}$ is 13.0 K, and the T$_{c,\,0}$
is 9.1 K as shown in Fig. \ref{fig3}(a). Since the magnetization signal is
dominated by the strong paramagnetic contribution from the oxide
substrates in low temperature, we subtract the magnetization of
the film from that of the film on MgO substrate, as given in Fig. \ref{fig3}(b),(c).
The superconducting volume is 22$\,$\% at 2 K, which is much higher
than FeSe \cite{HsuPNAS} and close to Fe$_{1+\delta}$Se$_{0.5}$Te$_{0.5}$ \cite{Sales}. Almost all
the film samples show T$_{c,\,onset}$ of 13.0 K, identified by the
resistance measurement and the dc magnetization measurement. This
value is much higher than that in FeSe bulk \cite{HsuPNAS,WuMKphysicaC} and
thin film \cite{HanY,WuMKphysicaC,WuMKprl} samples. The highest T$_{c,0}$
observed in resistance measurement is 10.6 K, for a 90 nm thick
film on SrTiO$_3$.
Further studies should pay attentions to the possible inhomogeneity inside the films which may contribute to the non-bulk superconductivity and relatively large $\Delta$T$_{c}$.

\begin{figure}
\includegraphics[angle=0,width=75mm]{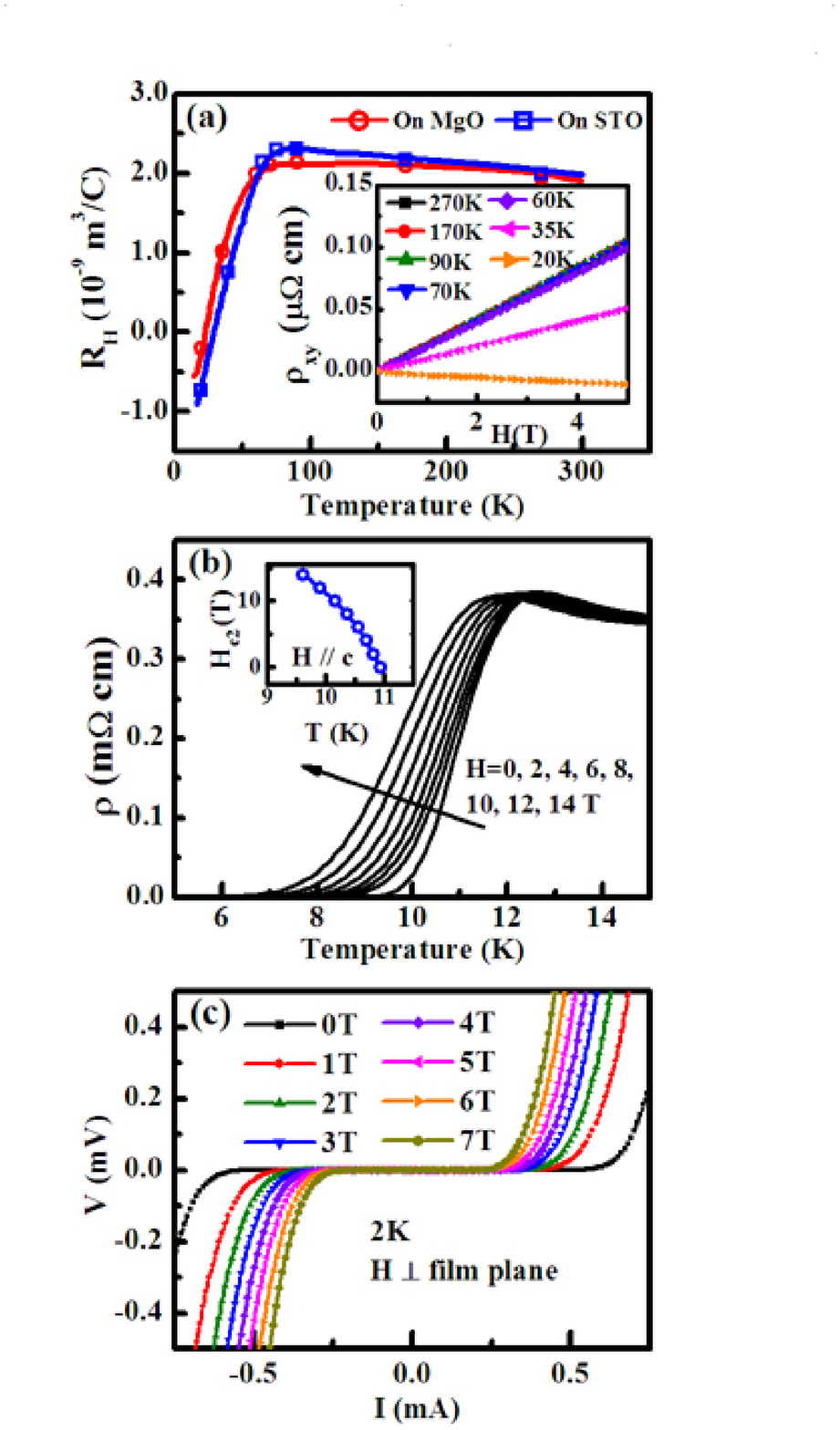}%
\caption{(color). Transport properties of our FeTe thin films. (a) Temperature
dependencies of the Hall coefficients of 90 nm thick FeTe films on
MgO and on SrTiO$_{3}$ measured at fixed magnetic field 5 T by scanning
the temperature. Inset gives the Hall resistivity as a function of
applied magnetic field measured at fixed temperatures for the film
on MgO. Results from temperature scan and field scan match well to
each other. A bar of $300\times 900\,\mu$m$^{2}$ was etched. (b) Influence of
magnetic field on superconductivity for a 90 nm FeTe film on SrTiO$_{3}$.
Inset shows upper critical fields deduced from the mid-point
transition temperatures with H$_{c2}$(0) extrapolated to be 123.0 T. A
microbridge of $10\times 100\,\mu$m$^{2}$ was etched. (c) Current versus voltage
measurements for the microbridge given in (b) performed at 2 K under
magnetic field up to 7 T. Critical current densities at 2 K under 0
T and 7 T are $6.7\times 10^{4}\,$A$\,$cm$^{-2}$ and $3.0\times 10^{4}\,$A$\,$cm$^{-2}$, respectively.
Critical currents were read at $1\times 10^{-5}\,$V criteria. \label{fig4}}
\end{figure}

In non-superconducting FeTe bulk samples, a first-order magnetic and
structural phase transition occurs at $\sim70$ K accompanied by the
anomalies in resistivity (Fig. \ref{fig3}(a)), magnetic susceptibility, and
Hall coefficient \cite{LiSLprb,BaoW,ChenGF}. Obviously this transition is broadened,
with maxima or dramatic drop starting at 85.7 K (Fig. \ref{fig3}(a)), 120.2 K
(Fig. \ref{fig3}(b)), and 79.3 K / 86.2 K (Fig. \ref{fig4}(a)) for the resisitivity,
susceptibility, and Hall coefficient, respectively. Furthermore, the
influences of the magnetic field on superconductivity were
investigated (Fig. \ref{fig4}(b),(c)). The upper critical field H$_{c2}$(0) estimated
\cite{HanY} is 123.0 T, much higher than that of FeSe \cite{HanY,HsuPNAS}, and
comparable to those of iron pnictides \cite{YuanHQ}. The current carrying
capacities, i.e., the critical current densities J$_{c}$(2K, 0T) and
J$_{c}$(2K, 7T), are $6.7\times 10^{4}$ A$\,$cm$^{-2}$ and $3.0\times 10^{4}\,$A$\,$cm$^{-2}$,
respectively. This suggests that the FeTe films may serve as a good
supercurrent carrier under certain magnetic field.

The tetragonal FeTe is in the spotlight because of its
theoretical significances: (1) There are increasing consensus that
superconductivity in the iron pnictides comes from doping induced
suppression of the spin-density-wave (SDW) ground state \cite{Norman}, while
if or not such picture applies fully to the iron chalcogenides is
still under debate \cite{Balatsky,XiaY,HanMJ}. (2) Density functional study predicted a
stronger SDW and therefore a higher T$_{c}$ in FeTe than FeSe was
expected \cite{Subedi}. Such speculation was growing after 37 K
superconductivity was reached under high pressure in FeSe \cite{Medvedev} and
application of pressure was found to enhance the spin fluctuations
\cite{ImaiT}. With superconducting FeTe films available, whether or not the
SDW exists \cite{Balatsky,XiaY,HanMJ} but only being softened as our experiments have
indicated will provide testimony to the ongoing debate upon
mechanism of iron pnictides and chalcogenides.

In summary, the onset of superconductivity of 13 K has been introduced
by interfacial stress, and more specifically by the tensile stress,
into the tetragonal non-superconducting FeTe compound.
We noticed that this can only be realized via stretching the specimen
rather than compressing, which can be regarded as a ``negative'' hydrostatic pressure.
FeTe has been regarded as touchstone of several appealing mechanism proposals.
The new findings, the softening of the first-order
transition and the increment of bond angles,
surely input more ingredients (or say, more
constraints) for further theoretical studies on superconducting mechanism.

\begin{acknowledgments}
This work is supported by the MOST of China (2006CB921107 and
2009CB320305) and the NSFC (10774165). L.X.C. and J.L.Y. are also
supported by the BRJH of the Chinese Academy of Sciences.
\end{acknowledgments}


\end{document}